\documentclass[twocolumn,floatfix,prl,showpacs,superscriptaddress]{revtex4}

\usepackage{float}
\usepackage{graphicx}
\usepackage{bm}
\usepackage{amsmath}
\usepackage{amssymb}
\usepackage{bbold}
\usepackage{xcolor}

\newcommand{\beq}{\begin{eqnarray}}
\newcommand{\eeq}{\end{eqnarray}}
\newcommand{\beqq}{\begin{eqnarray*}}
\newcommand{\eeqq}{\end{eqnarray*}}

\begin{document}

\title{Fractons from Partons}

\author{Timothy H. Hsieh}
\affiliation{Kavli Institute for Theoretical Physics, University of
California, Santa Barbara, CA 93106, USA}
\author{G\'abor B. Hal\'asz}
\affiliation{Kavli Institute for Theoretical Physics, University of
California, Santa Barbara, CA 93106, USA}

%\pacs{71.10.Fd, 71.10.Pm, 73.20.-r, 11.30.Pb}

\begin{abstract}

Fracton topological phases host fractionalized excitations that are
either completely immobile or only mobile along certain lines or
planes.  We demonstrate how such phases can be understood in terms
of two fundamentally different types of parton constructions, in
which physical degrees of freedom are decomposed into clusters of
``parton'' degrees of freedom subject to emergent gauge constraints.
First, we employ non-interacting partons subject to multiple
overlapping constraints to describe a fermionic fracton model.
Second, we demonstrate how {\it interacting} partons can be used to
develop new models of bosonic fracton phases, both with string and
membrane logical operators (type-I fracton phases) and with fractal
logical operators (type-II fracton phases). In particular, we find a
new type-II model which saturates a bound on its information storage
capacity. Our parton approach is generic beyond exactly solvable
models and provides a variational route to realizing fracton phases
in more physically realistic systems.

\end{abstract}

\maketitle

Topologically ordered phases in three dimensions have presented a
wealth of surprising phenomena, challenging conventional paradigms
such as effective field theory and the notion of the thermodynamic
limit itself. In particular, ``fracton phases'' \cite{chamon, blt,
haahcodes, yoshida, majcube, pretko} are characterized by
fractionalized excitations that are either completely immobile or
mobile only along lines or planes, despite translation symmetry.
More precisely, moving such excitations requires creating even more
such excitations, and this energetic barrier against motion presents
not only an exciting alternative to (disorder-driven) many-body
localization \cite{localization, glass} but also a marginally stable
quantum memory at finite temperature \cite{logmemory-1,
logmemory-2}.

Much progress has been made in both understanding such fracton
phases and developing new models. For example, fracton phases have
been related to gauged classical systems with subsystem symmetry
\cite{subsystem, williamson} and coupled-layer constructions
\cite{hermele, vijay}. Nonetheless, there remain many open questions
involving possible field-theoretical descriptions and phase
transitions out of fracton phases. Moreover, while there have been
many new fracton models with string and membrane logical operators
(``type-I'' fracton models), the number of models with fractal
logical operators (``type-II'' fracton models) has thus far been
limited \cite{haahcodes, yoshida}.

The goal of our work is to construct and understand fracton phases
by decomposing physical degrees of freedom into clusters of
``partons''. The parton approach has proven to be extremely valuable
in illuminating the physics of interacting topological phases. In
some remarkable cases, partons furnish exact solutions of
spin-liquid models \cite{kitaev, wen-1}, and they correspond
directly to the deconfined excitations of the physical system. More
generally, partons have provided useful variational wavefunctions
for otherwise intractable spin systems.

In this work, we demonstrate how parton constructions can be used to
describe, develop, and analyze fracton models. We first provide an
explicit parton construction of a fermionic fracton model, in which
the partons are non-interacting but are subject to multiple
overlapping gauge constraints. We also construct a new framework
involving {\it interacting} partons to develop new models of bosonic
fracton phases. We illustrate this construction with exact parton
descriptions of two new fracton models, one with string and membrane
and one with fractal logical operators. The model with fractal
logical operators is beyond the original Haah codes and notably
saturates a bound on the number of encoded qubits \cite{haahcodes}.
Our parton language provides a new perspective on fracton phases,
potentially furnishes a route to even more exotic phases, and may
suggest more physically realistic fracton models via a variational
approach.

\section{General Considerations}

We begin by reviewing Kitaev's parton construction of spins in terms
of Majorana fermions \cite{kitaev}, which serves as a basis for our
more involved parton constructions. For a system of spin-half
degrees of freedom, one can represent each spin $\sigma$ by four
Majorana fermions (partons) $\gamma_{1,2,3,4}$ as $\sigma^x = i
\gamma_1 \gamma_4$, $\sigma^y = i \gamma_2 \gamma_4$, and $\sigma^z
= i \gamma_3 \gamma_4$. One can then guess an approximate ground
state (i.e., a variational state) for the spin system by first
considering the ground state of a non-interacting (quadratic)
Hamiltonian for the partons. The key assumption behind such a guess
is that the partons are emergent quasiparticles whose behavior is
approximately governed by a non-interacting Hamiltonian.
Importantly, however, there are specific models where the parton
construction provides an exact solution to the spin system and
therefore no assumptions are necessary. Notable examples include the
Kitaev honeycomb model \cite{kitaev} and the Wen plaquette model
\cite{wen-1}.

Since the Hilbert space is enlarged in the parton representation,
the four partons of a given spin are subject to the constraint $G =
\gamma_1 \gamma_2 \gamma_3 \gamma_4 = 1$. The physical variational
state is then obtained from the ground state of the parton
Hamiltonian by a projection imposing this constraint for each spin.
From the partons' point of view, the constraint operators $G$ are
local gauge transformations (LGTs), and the resulting parton gauge
theory is crucial for understanding the physical state. Indeed, one
can classify spin-liquid phases by their parton variational states
\cite{wen-2} via the invariant gauge group (IGG): the subgroup of
the gauge group that leaves the parton Hamiltonian (and hence the
parton state) invariant. For example, the Kitaev honeycomb model and
the Wen plaquette model are both $\mathbb{Z}_2$ spin liquids
($\mathbb{Z}_2$ gauge theories) because their parton states are
invariant under a $\mathbb{Z}_2$ gauge group whose only non-trivial
element is the global gauge transformation (the product of all
LGTs).

Based on their ground-state degeneracies \cite{haahcodes} and their
logical operators \cite{yoshida}, fracton phases can be understood
as $\mathbb{Z}_2^N$ gauge theories, where $N$ is infinite in the
thermodynamic limit but has subextensive system-size scaling for a
finite system. In particular, for an $L \times L \times L$ system,
$N$ grows linearly with $L$ for type-I fracton phases and spikes at
particular values of $L$ for type-II fracton phases. Alternatively,
if one understands fracton phases as classical systems with gauged
subsystem symmetries \cite{subsystem, williamson}, one expects the
IGGs of the corresponding parton states to contain elements that are
products of LGTs along the appropriate subsystems, i.e., planes for
type-I phases and fractals for type-II phases. By considering how
these subsystems fit into an $L \times L \times L$ system, one can
then understand why the resulting IGGs should be $\mathbb{Z}_2^N$
with $N$ depending on $L$ as described above.

According to this understanding, Kitaev's parton construction with a
non-interacting Majorana Hamiltonian is insufficient to describe a
fracton phase because it necessarily gives rise to a simple
$\mathbb{Z}_2$ IGG. Since a two-Majorana term can only connect two
different spins via one Majorana from each spin, the Majorana
Hamiltonian can only be invariant under the product of the two
corresponding LGTs. For a connected system, it immediately follows
that the only non-trivial element of the IGG is then the product of
all LGTs. This obstacle motivates us to change Kitaev's parton
construction in two different ways. First, we consider a parton
construction where each constraint (i.e., each LGT) is substituted
with several overlapping constraints. Second, we consider
interacting parton Hamiltonians with four-Majorana terms that can
each connect four different spins.

\begin{figure}
\centering
\includegraphics[height=2.5in]{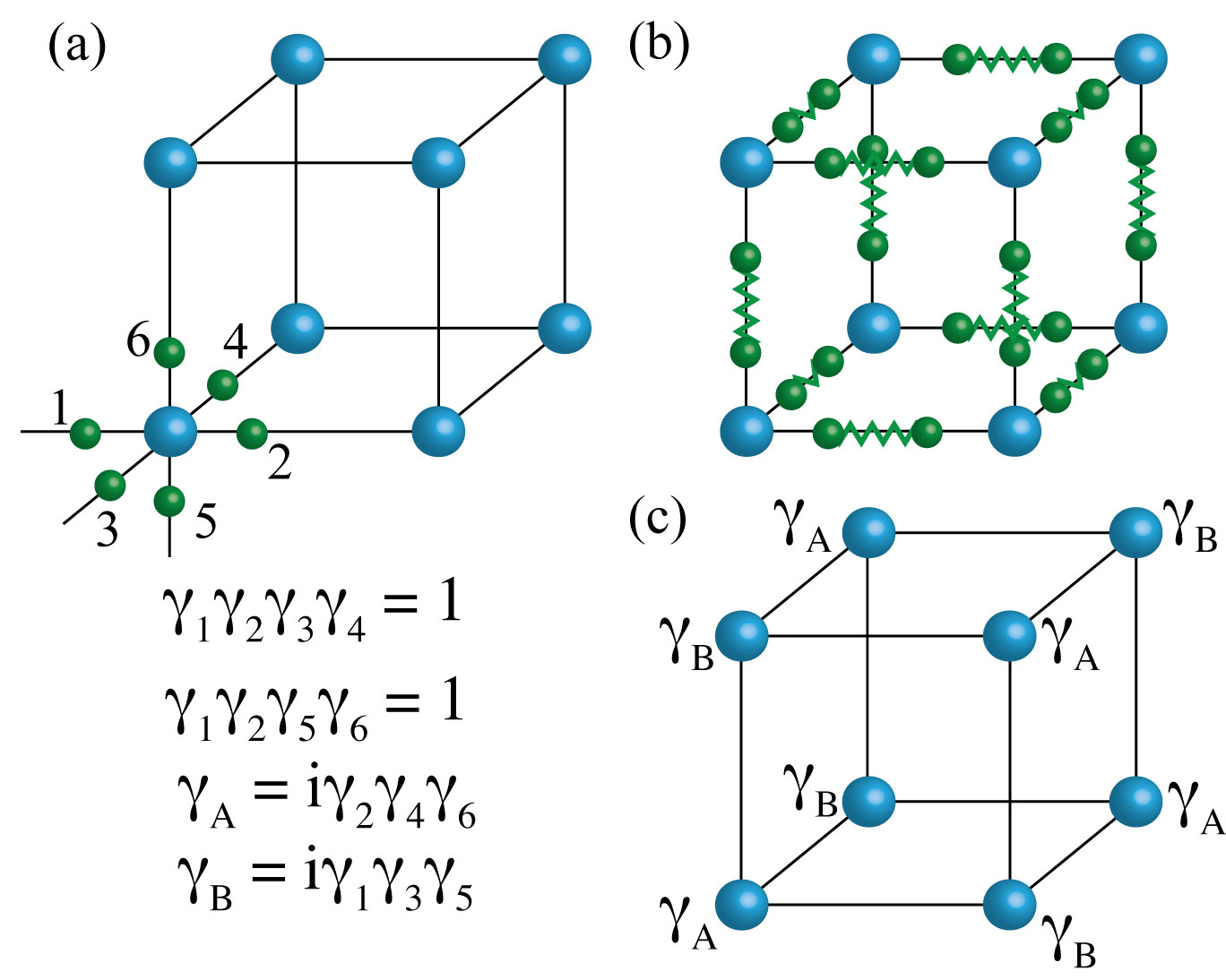}
\caption{{\bf Fermionic type-I fracton model from non-interacting
partons.} (a) Two physical Majorana fermions $\gamma_{A,B}$ at each
site (blue sphere) are decomposed into six Majorana partons
$\gamma_{1,...,6}$ (green spheres) subject to two constraints. (b)
Parton state specified by imposing $i\gamma_j\gamma_k = 1$ for each
link (green zigzag). (c) Eight-Majorana term of the parent
Hamiltonian whose ground state is the projected parton state.}
\end{figure}

\section{Fermionic Fracton Model}

We first consider a modified version of Kitaev's parton construction
where the physical degrees of freedom are not spins but two Majorana
fermions $\gamma_{A,B}$ at each site of a cubic lattice. These two
physical Majorana fermions are represented by six Majorana partons
$\gamma_{1,2,3,4,5,6}$ on the links adjacent to the site [see
Fig.~1(a)]. For example, we may take $\gamma_A = i \gamma_2 \gamma_4
\gamma_6$ and $\gamma_B = i \gamma_1 \gamma_3 \gamma_5$. To account
for the enlarged Hilbert space, we also impose two independent
constraints $G_{xy} = \gamma_1 \gamma_2 \gamma_3 \gamma_4 = 1$ and
$G_{xz} = \gamma_1 \gamma_2 \gamma_5 \gamma_6 = 1$, whose product is
a dependent third constraint $G_{yz} = \gamma_3 \gamma_4 \gamma_5
\gamma_6 = 1$. Note that these three gauge constraints are
directional in the sense that they only act (respectively) on
partons in the $xy$, $xz$, and $yz$ planes containing the given site
[see Fig.~1(a)].

The parton state is constructed by imposing the constraint $i
\gamma_j \gamma_k = 1$ for each pair of partons on the same link
[see Fig.~1(b)]. Since these constraints commute, the corresponding
parton Hamiltonian is simply the commuting-projector Hamiltonian
$-\sum_{\langle j, k \rangle} i \gamma_j \gamma_k$. For such a
parton state, the IGG has several non-trivial elements due to the
directionality of the gauge constraints. In particular, the product
of all $G_{xy}$ in any $xy$ plane, the product of all $G_{xz}$ in
any $xz$ plane, and the product of all $G_{yz}$ in any $yz$ plane
each leave the parton state invariant.  However, the product of all such ``planar'' IGG elements is trivial.  For an $L \times L \times L$
system, the IGG is then $\mathbb{Z}_2^{3L-1}$, which is indicative of
a type-I fracton phase.

What parent Hamiltonian in terms of the physical Majorana fermions
could have the projected parton state as its ground state? The
topological bootstrap introduced by one of us \cite{bootstrap}
provides one route for deriving such a parent Hamiltonian. One seeks
the minimal Hamiltonian in terms of the physical degrees of freedom,
which, when written in terms of the partons, commutes with the
parton Hamiltonian. Using this technique, we find that the parent
Hamiltonian involves a product of eight physical Majorana fermions
for each cube [see Fig.~1(c)]. In the parton representation, the
eight-Majorana term for each cube is then the product of the twelve
terms $i \gamma_j \gamma_k$ on the twelve links surrounding the cube
\cite{sm}.

This model is equivalent to two copies of the Majorana cube model
(MCM), which was introduced in Ref.~\onlinecite{majcube} as a type-I
fracton model. The MCM has only one flavor of Majorana fermion
$\gamma_A$ at each site and involves interactions for only half of
the cubes (i.e., a subset of cubes which are either disjoint or
intersect at an edge). Our model can then be reproduced by taking
one more copy of the MCM on the complementary subset of cubes
involving a different flavor of Majorana fermion $\gamma_B$, and
using a unitary transformation $\gamma_A \leftrightarrow \gamma_B$
on one sublattice of the (bipartite) cubic lattice.

Since our model is two copies of the MCM, it clearly captures a
type-I fracton phase as well. Indeed, one can enumerate all
characteristic fractional excitations of such a type-I fracton phase
in our model \cite{majcube}. First, a string of alternating Majorana
flavors $\ldots \gamma_A \gamma_B \ldots$ along the $x$ direction
creates two pairs of excitations at its endpoints that are mobile
along the $x$ direction only. Second, a double string where two
complementary strings $\ldots \gamma_A \gamma_B \ldots$ and $\ldots
\gamma_B \gamma_A \ldots$ are displaced in the $z$ direction creates
two pairs of excitations at its endpoints that are mobile in the
$xy$ plane. Third, a rectangular checkerboard of $\gamma_A$ and
$\gamma_B$ operators creates four excitations at its corners that
are completely immobile.

Our parton construction is suggestive of coupling two-dimensional
topologically ordered stacks \cite{hermele, vijay}. Indeed, if one
imposed only one gauge constraint $G_{xy}$, $G_{xz}$, or $G_{yz}$ at
each site, the model would consist of decoupled stacks of Wen
plaquette models \cite{wen-1} with decoupled Majorana dimers in
between. It is the presence of all three directional gauge
constraints that produces a type-I fracton model. However, these
constraints remarkably conspire to produce a {\it fermionic} fracton
model.

\section{Bosonic Fractons from Interacting Partons}

While it is possible to describe a bosonic fracton phase by
non-interacting partons, a more natural choice for us, especially
for describing type-II phases with fractal structures, is to
consider a different construction involving {\it interacting}
partons. Naively, this may not seem useful because interacting
partons are in general as difficult to analyze as the (interacting)
physical degrees of freedom. However, we focus on parton variational
states that are ground states of interacting yet exactly solvable
commuting-projector Hamiltonians.

\begin{figure}
\centering
\includegraphics[height=3in]{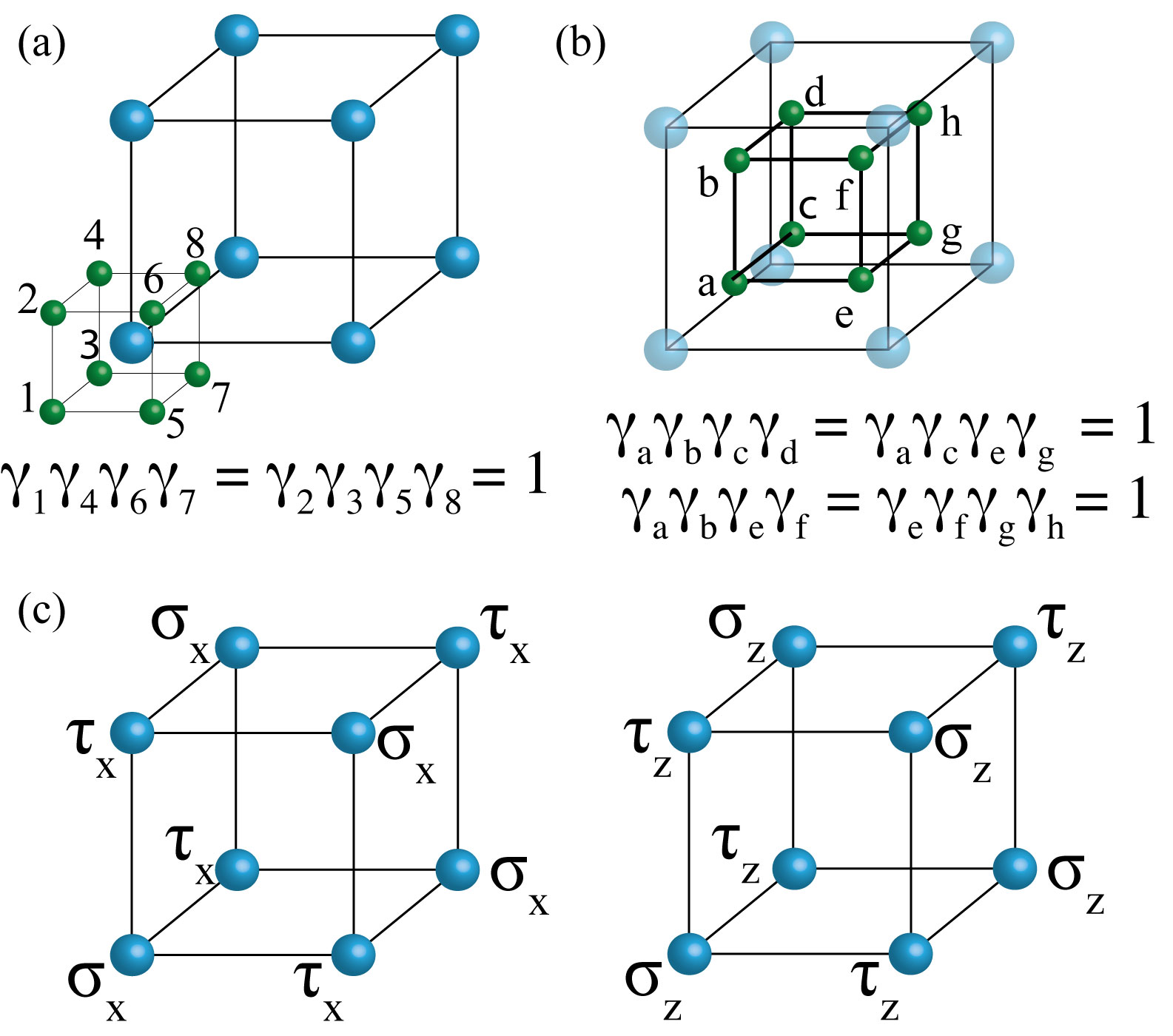}
\caption{{\bf Bosonic type-I fracton model from interacting
partons.} (a) Two spin-one-halves at each site (blue sphere) are
decomposed into eight Majorana partons (green spheres) subject to
two constraints. (b) Parton state specified by four constraints for
each parton cube. (c) Independent eight-spin interactions of the
parent Hamiltonian.}
\end{figure}

\subsection{Type-I Fracton Model}

We first present an explicit example of such a construction that
yields a new type-I fracton model with string and membrane logical
operators. The physical degrees of freedom are two spin-one-halves
$\sigma$ and $\tau$ at each site of a cubic lattice. Each spin is
represented by four Majorana partons and is subject to a single
constraint, as in Kitaev's original construction \cite{kitaev}. In
particular, the two spins at any given site are associated with the
two tetrahedra formed by the eight Majorana partons $\gamma_{1,
\ldots, 8}$ surrounding the site [see Fig.~2(a)]. The two
constraints are then $G_{\sigma} = \gamma_1 \gamma_4 \gamma_6
\gamma_7$ and $G_{\tau} = \gamma_2 \gamma_3 \gamma_5 \gamma_8$,
while the spin components are $\sigma^x = i \gamma_1 \gamma_4$,
$\sigma^y = i \gamma_1 \gamma_6$, $\sigma^z = i \gamma_1 \gamma_7$,
$\tau^x = i \gamma_5 \gamma_8$, $\tau^y = i \gamma_3 \gamma_8$, and
$\tau^z = i \gamma_2 \gamma_8$.

The parton state is constructed as follows. Each cube of the
original cubic lattice contains eight Majorana partons from spins at
eight different sites [see Fig.~2(b)]. For each such parton cube
consisting of eight Majorana partons, we impose a Hamiltonian
involving a four-Majorana term at each face of the cube. Since it
has four independent (and two dependent) commuting terms, this
parton Hamiltonian gives rise to a unique local ground state for
each parton cube \cite{susy}. The parton state is then simply the
direct product of these local ground states.

For such a parton state, the IGG has several non-trivial elements.
First, the product of all $G_{\sigma} G_{\tau}$ in any $xy$, $xz$,
or $yz$ plane leaves the parton state invariant. Second, if such a
checkerboard pattern is consistent with the system size, the product
of all $G_{\sigma}$ in one sublattice and all $G_{\tau}$ in the
other sublattice of any $xy$, $xz$, or $yz$ plane also leaves the
parton state invariant. However, these elements of the IGG are not
all independent. In fact, for an $L \times L \times L$ system, a
detailed analysis shows that the IGG is $\mathbb{Z}_2^{3L-1}$ if $L$
is odd and $\mathbb{Z}_2^{6L-4}$ if $L$ is even \cite{sm}.

Once again, the topological bootstrap can be used to obtain a parent
spin Hamiltonian whose ground state is the projected parton state.
Using this technique, we find that the parent spin Hamiltonian has
two independent eight-spin interactions for each cube [see
Fig.~2(c)]. In the parton representation, each eight-spin term is
then the product of four four-Majorana terms corresponding to four
faces of neighboring parton cubes \cite{sm}.

The system-size dependence of the IGG indicates that this model
captures a type-I fracton phase. Indeed, the fractional excitations
of this bosonic model are supported by string and membrane logical
operators, which can be obtained from those of our fermionic model
via the substitutions (i) $\gamma_A \rightarrow \sigma^x$ and
$\gamma_B \rightarrow \tau^x$ and (ii) $\gamma_A \rightarrow
\sigma^z$ and $\gamma_B \rightarrow \tau^z$. For example, a string
of alternating spin types $\ldots \sigma^x \tau^x \ldots$ or $\ldots
\sigma^z \tau^z \ldots$ along the $x$, $y$, or $z$ direction creates
two pairs of excitations at its endpoints that are mobile along the
string direction only. Also, as expected for a type-I fracton model,
the number of encoded qubits scales with the linear dimension of the
system.

\subsection{Type-II Fracton Model}

We now demonstrate that our interacting parton construction can also
yield a type-II fracton model with fractal logical operators. In
fact, we derive a spin model beyond the original Haah codes and find
that it saturates a bound on the number of encoded qubits
\cite{haahcodes}. Remarkably, our type-II construction is obtained
from the type-I construction above by simply changing which set of
eight Majorana partons interact with each other. Indeed, we use the
same representation of spins in terms of Majorana partons [see
Fig.~3(a)], but we choose a different unit cell of eight Majorana
partons for the four-Majorana terms in our parton Hamiltonian [see
Fig.~3(b)]. The parton state is again the direct product of the
unique ground states for these eight-Majorana unit cells.

For such a parton state, the IGG is $\mathbb{Z}_2^N$, where $N$ has
a peculiar dependence on the system size. For an $L \times L \times
L$ system, $N$ is only $2$ if $L$ is a generic odd number, while it
reaches $2L$ if $L = 2^n$ \cite{sm}. In the former case, the only
non-trivial elements of the IGG are global ones: the products of all
$G_{\sigma}$ and of all $G_{\tau}$. In the latter case, however,
there are further non-trivial elements corresponding to products of
$G_{\sigma}$ and $G_{\tau}$ along fractal structures. Since the
scaling ratio of each fractal structure is $2$, it only fits into
the system if $L = 2^n \ell$, where $\ell$ is the size of its base
unit. For our fractal structures, the two smallest base units
correspond to $\ell_1 = 1$ and $\ell_2 = 15$.

The topological bootstrap can again be used to obtain a parent spin
Hamiltonian whose ground state is the projected parton state. Once
again, we find that the parent spin Hamiltonian has two independent
eight-spin interactions for each cube [see Fig.~3(c)] and that each
eight-spin term is then the product of four four-Majorana terms in
the parton representation \cite{footnote-1}.

\begin{figure}
\centering
\includegraphics[height=3in]{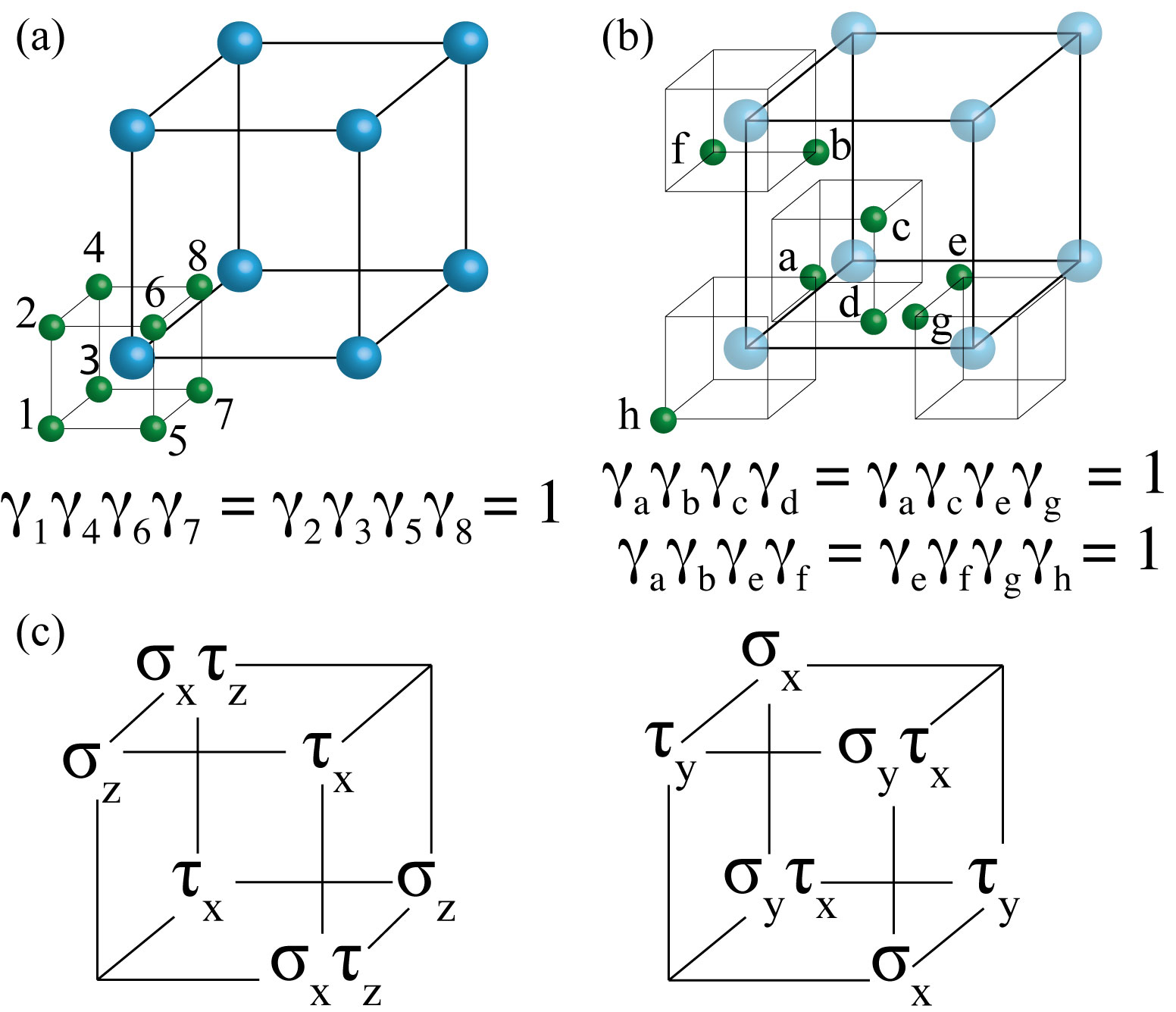}
\caption{{\bf Bosonic type-II fracton model from interacting
partons.} (a) Two spin-one-halves at each site (blue sphere) are
decomposed into eight Majorana partons (green spheres) subject to
two constraints. (b) Parton state specified by four constraints for
each eight-Majorana unit cell. (c) Independent eight-spin
interactions of the parent Hamiltonian.}
\end{figure}

The system-size dependence of the IGG indicates that this model
captures a type-II fracton phase. Indeed, the fractional excitations
of this model are supported by fractal logical operators. For example, one such fractal operator is
constructed iteratively as follows. First, the local operator
$\sigma^x \tau^y$ anticommutes with six interaction terms and thus
creates six excitations on the dual lattice [see Fig.~4(a)]. Next,
by taking the product of six $\sigma^x \tau^y$ operators at the six
points shown in Fig.~4(b), the resulting set of excitations is
identical in shape but is rescaled by a factor of $2$ with respect
to the original one. This iterative procedure generates a fractal
operator of size $2^n$ with six immobile excitations at its corners.

The model presented in Fig.~3(c) has several interesting features
with respect to previously known type-II fracton models
\cite{haahcodes, yoshida}. As expected for any such model, the
number of encoded qubits for an $L \times L \times L$ system has
large spikes for $L = 2^n$. However, unlike any of the original Haah
codes, our model can encode $4L$ qubits for $L = 2^n$, thereby
saturating the upper bound for the number of encoded qubits in a
type-II fracton system with two qubits per site and interactions supported on single cubes \cite{haahcodes}.
Furthermore, our model is a non-CSS code as each interaction
involves both $x$-type and $z$-type spin operators, and it is
therefore not clear how to realize it by gauging a classical spin
model \cite{subsystem, williamson}.  Finally, to our knowledge, there are no string logical operators in this
model, although rigorously proving this claim is challenging as
techniques used in Refs.~\cite{haahcodes} and \cite{yoshida} are not
directly applicable.

\begin{figure}
\centering
\includegraphics[height=1.7in]{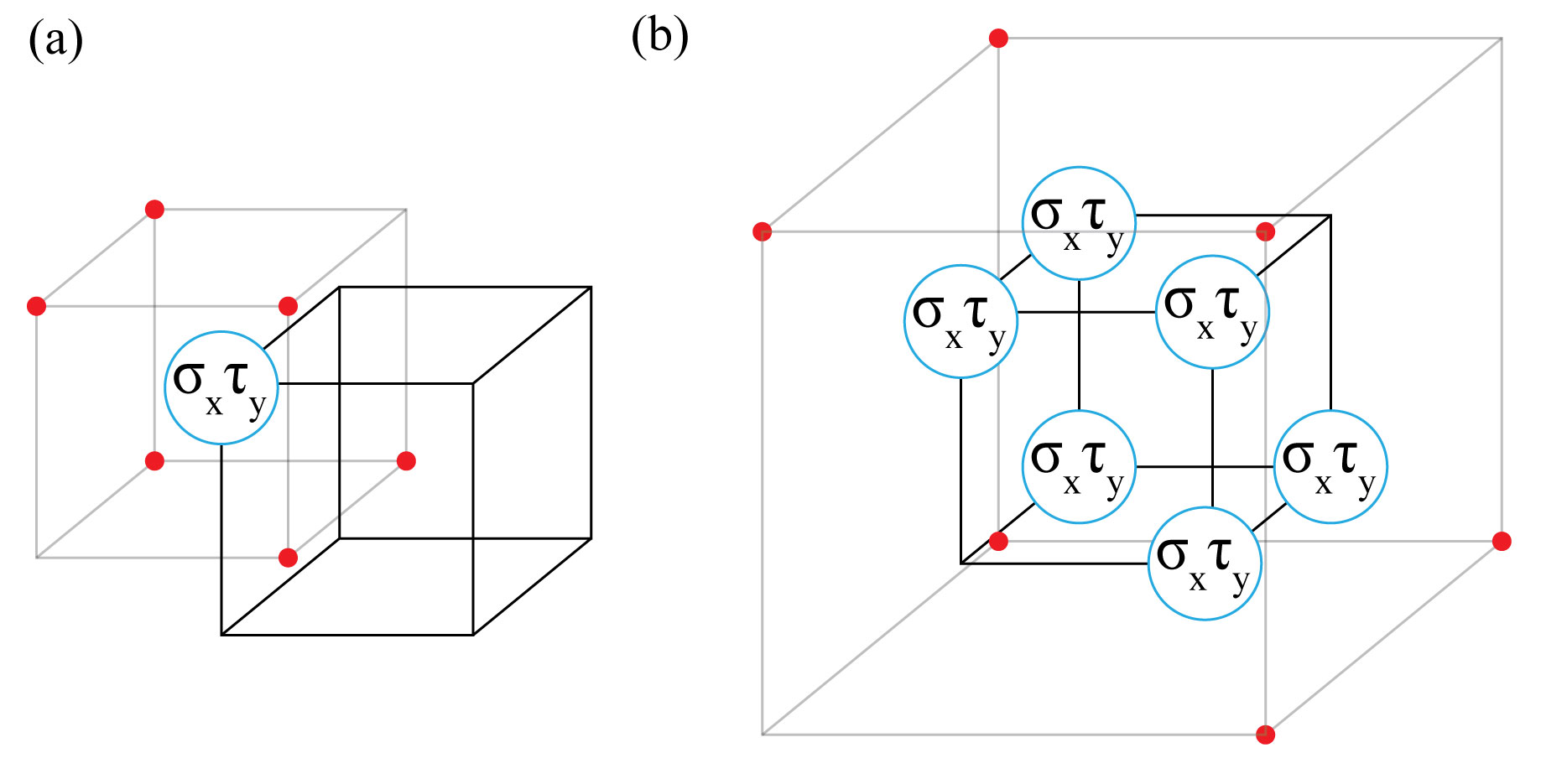}
\caption{{\bf Fractal structure in our type-II fracton model.} (a)
For the model in Fig.~3(c), the single-site operator $\sigma^x
\tau^y$ creates six excitations (red dots). (b) Six such operators
in the given configuration create six defects with the same shape as
in (a) but with doubled linear dimension.}
\end{figure}

\section{Summary and Discussion}

We have provided two different ways of describing and developing new
fracton models by means of parton constructions. The first method
uses non-interacting partons with multiple overlapping gauge
constraints at each site and was used to obtain a fermionic type-I
model, while the second method uses interacting partons governed by
a commuting-projector Hamiltonian and was used to obtain bosonic
type-I and type-II models. The new type-II model is particularly
interesting because (i) it is the first non-CSS code involving qubits that captures a
type-II fracton phase \cite{kim} and (ii) it saturates a bound on the maximal
number of encoded qubits for particular system sizes.

In addition to providing exactly solvable fracton models, our parton
approach may also enable a variational treatment of more realistic
models that are not exactly solvable but are suspected to capture
fracton phases. In fact, there are many variants of our interacting
parton constructions that do not give rise to exactly solvable
models. Nevertheless, based on the system-size dependence of their
IGGs and the structures of their IGG elements, one can immediately
deduce whether they correspond to an ordinary topological phase, a
type-I fracton phase, or a type-II fracton phase. Furthermore, in
the case of type-II fracton phases, one expects a direct
correspondence between the fractal structures of the IGG elements
and those of the logical operators.

There are many directions for extending and utilizing this parton
approach, for example, in investigating phase transitions between
fracton and other phases. Moreover, it would be interesting to apply
the formalism of projective symmetry group, which has been useful
for classifying conventional topological order, to fracton phases in
the presence of symmetries.

\begin{acknowledgements}
{\it Acknowledgements:} We especially thank Leon Balents
and Chao-Ming Jian, and we also thank Tarun Grover, Yuan-Ming Lu, and Beni Yoshida
for useful discussions. Both authors are supported by a fellowship
from the Gordon and Betty Moore Foundation (Grant No.~4304).
\end{acknowledgements}

%%%%%%%%%%%%%%%%%%%%%%%%%%%%%%%%%%%%%%%%%%%%%%%%%

\clearpage

\begin{widetext}

\subsection{\large Supplementary Material}

\subsection{Invariant gauge groups and ground-state degeneracies}

For our interacting parton constructions and the corresponding
exactly solvable spin models, one is interested in the invariant
gauge group (IGG) of the parton construction and the ground-state
degeneracy of the spin model as a function of the system size. In
this section, we demonstrate that these quantities can be evaluated
straightforwardly by means of standard $\mathbb{Z}_2$ linear
algebra.

We first consider the parton constructions. For a given system size,
we assume that there are $N_{\gamma}$ Majorana partons, $N_H$
interaction terms in the parton Hamiltonian, and $N_G$ independent
gauge constraints. Each interaction term $H$ (gauge constraint $G$)
is a product of Majorana partons and it can thus be represented by
an $N_{\gamma}$-component vector $\mathbf{h}$ ($\mathbf{g}$) of
$\mathbb{Z}_2$ elements such that each element is $1$ if the product
contains the corresponding Majorana parton and $0$ if it does not.
Furthermore, we may include all interaction terms in the $N_H \times
N_{\gamma}$ matrix $\mathbf{H}$ and all gauge constraints in the
$N_G \times N_{\gamma}$ matrix $\mathbf{G}$. In general, an
interaction term and a gauge constraint either commute or
anticommute:
\begin{eqnarray}
[H, G] = 0 \quad &\Longleftrightarrow& \quad \mathbf{h} \cdot
\mathbf{g} = 0 \quad (\textrm{mod } 2), \nonumber \\
\{H, G\} = 0 \quad &\Longleftrightarrow& \quad \mathbf{h} \cdot
\mathbf{g} = 1 \quad (\textrm{mod } 2). \nonumber
\end{eqnarray}
By definition, each element of the IGG is an appropriate product of
gauge constraints $G$ that commutes with all the interaction terms
$H$. The number of independent IGG elements is then
\begin{eqnarray}
N = N_G - \textrm{rank} \left[ \mathbf{H} \cdot \mathbf{G}^T
\right], \nonumber
\end{eqnarray}
while the IGG elements themselves are contained in $\textrm{ker}
[\mathbf{H} \cdot \mathbf{G}^T]$. Note that the rank and the kernel
(null space) both must be taken modulo $2$. Since all elements are
$\mathbb{Z}_2$, the IGG is given by $\mathbb{Z}_2^N$.

We next consider the corresponding spin models. For a given system
size, we assume that there are $N_{\sigma}$ spin-one-half degrees of
freedom and there are $N_{\tilde{H}}$ commuting interaction terms in
the spin Hamiltonian. Each interaction term $\tilde{H}$ is a product
of spin operators $\sigma^x$, $\sigma^y$, and $\sigma^z \propto
\sigma^x \sigma^y$ and it can thus be represented by a
$2N_{\sigma}$-component vector $\tilde{\mathbf{h}}$ of
$\mathbb{Z}_2$ elements such that each pair of elements is $\{ 0,0
\}$ if the interaction term does not act at the corresponding spin,
while it is $\{ 0,1 \}$, $\{ 1,0 \}$, and $\{ 1,1 \}$ if the
interaction term acts at the corresponding spin by $\sigma^x$,
$\sigma^y$, and $\sigma^z$ operators, respectively. Furthermore, we
may include all interaction terms in the $N_{\tilde{H}} \times
2N_{\sigma}$ matrix $\tilde{\mathbf{H}}$. In general, the
interaction terms are not all independent and a subset $\{k\}$ of
them may satisfy
\begin{eqnarray}
\prod_{\{k\}} \tilde{H}_k \propto 1 \quad &\Longleftrightarrow&
\quad \sum_{\{k\}} \tilde{\mathbf{h}}_k = 0 \quad (\textrm{mod } 2).
\nonumber
\end{eqnarray}
Since the Hilbert space contains $N_{\sigma}$ effective qubits, and
each independent (commuting) interaction term specifies one
effective qubit, the actual number of qubits encoded in global
degrees of freedom is
\begin{eqnarray}
\tilde{N} = N_{\sigma} - \textrm{rank} [\tilde{\mathbf{H}}].
\nonumber
\end{eqnarray}
Once again, the rank must be taken modulo $2$. For $\tilde{N}$ such
qubits encoded in global degrees of freedom, the ground-state
degeneracy of the spin model is then given by $2^{\tilde{N}}$.

For our interacting parton constructions, the number $N$ of
$\mathbb{Z}_2$ factors in the IGG is given in Table I, while for the
corresponding spin models, the number $\tilde{N}$ of encoded qubits
is given in Table II. For the purpose of benchmarking, we also
include the Wen plaquette model and its parton construction. While
$N$ and $\tilde{N}$ might not have identical system-size dependence
and, in particular, they might have different even-odd oscillations,
they follow the same qualitative behavior: they are both
approximately constant for the Wen plaquette model, scale with the
linear system dimension for our type-I fracton model, and spike at
particular system sizes for our type-II fracton model.

\begin{table}[h]
\begin{tabular*}{0.7\textwidth}{@{\extracolsep{\fill}} c | c | c | c | c | c | c | c | c | c | c | c | c | c | c | c | c }
\hline \hline
System size ($L$)     &  1 &  2 &  3 &  4 &  5 &  6 &  7 &  8 &  9 & 10 & 11 & 12 & 13 & 14 & 15 & 16 \\
\hline
Wen plaquette model   &  1 &  1 &  1 &  1 &  1 &  1 &  1 &  1 &  1 &  1 &  1 &  1 &  1 &  1 &  1 &  1 \\
\hline
Type-I fracton model  &  2 &  8 &  8 & 20 & 14 & 32 & 20 & 44 & 26 & 56 & 32 & 68 & 38 & 80 & 44 & 92 \\
\hline
Type-II fracton model &  2 &  4 &  2 &  8 &  2 &  4 &  2 & 16 &  2 &  4 &  2 &  8 &  2 &  4 & 26 & 32 \\
\hline  \hline
\end{tabular*}
\caption{Number $N$ of $\mathbb{Z}_2$ factors in the IGG for an $L
\times L$ system in the case of the Wen plaquette model and for an
$L \times L \times L$ system in the case of our bosonic fracton
models. \label{table-1}}
\end{table}

\begin{table}[h]
\begin{tabular*}{0.7\textwidth}{@{\extracolsep{\fill}} c | c | c | c | c | c | c | c | c | c | c | c | c | c | c | c | c }
\hline \hline
System size ($L$)     &  1 &  2 &  3 &  4 &  5 &  6 &  7 &  8 &  9 &  10 & 11 &  12 & 13 &  14 & 15 &  16 \\
\hline
Wen plaquette model   &  1 &  2 &  1 &  2 &  1 &  2 &  1 &  2 &  1 &   2 &  1 &   2 &  1 &   2 &  1 &   2 \\
\hline
Type-I fracton model  &  2 & 12 & 14 & 36 & 26 & 60 & 38 & 84 & 50 & 108 & 62 & 132 & 74 & 156 & 86 & 180 \\
\hline
Type-II fracton model &  2 &  8 &  2 & 16 &  2 &  8 &  2 & 32 &  2 &   8 &  2 &  16 &  2 &   8 & 50 &  64 \\
\hline  \hline
\end{tabular*}
\caption{Number $\tilde{N}$ of encoded qubits for an $L \times L$
system in the case of the Wen plaquette model and for an $L \times L
\times L$ system in the case of our bosonic fracton models.
\label{table-2}}
\end{table}

\subsection{Parent Hamiltonians for Projected Parton States}

The topological bootstrap \cite{bootstrap} provides a simple
heuristic to obtain a parent Hamiltonian for a projected parton
state. We briefly review the bootstrap construction, but ultimately
only the heuristic is necessary. In this section, we first review
how both the bootstrap and the heuristic apply to the Wen plaquette
model, and then we apply the heuristic to the parton constructions
of the fracton models considered in the main text.

The Wen plaquette model consists of spin-1/2s at the sites of a
square lattice, with plaquette interactions given by \beq H = \sum_l
\sigma^y_l \sigma^x_{l+x} \sigma^y_{l+x+y} \sigma^x_{l+y}. \nonumber
\eeq Following Kitaev \cite{kitaev} and Wen \cite{wen-1}, each spin
is represented by four Majorana fermions $\gamma_{1,2,3,4}$ on the
links adjacent to the spin site [see Fig.~5(a)] that are subject to
the constraint $\gamma_1 \gamma_2 \gamma_3 \gamma_4 = 1$. The parton
state is defined as the ground state of the Hamiltonian
$-\sum_{\langle j, k \rangle} i \gamma_j \gamma_k$, where $\langle
j, k \rangle$ label Majoranas on the same link [see Fig.~5(b)].

\begin{figure}
\centering
\includegraphics[height=1.5in]{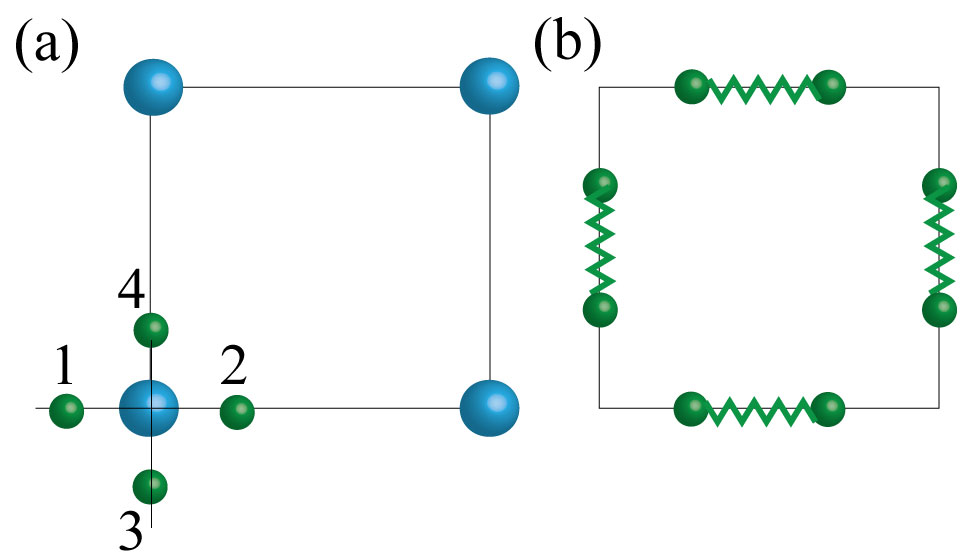}
\caption{(a) The physical spin-1/2 at each site (blue sphere) is
decomposed into four Majorana partons $\gamma_{1,2,3,4}$ (green
spheres) subject to the constraint $\gamma_1 \gamma_2 \gamma_3
\gamma_4 = 1$. (b) The parton state for the Wen plaquette model is
the product state of dimers formed by pairs of Majoranas on each
link (green zigzag).}
\end{figure}

Given this parton state, how would one obtain the parent spin
Hamiltonian without knowing about the Wen plaquette model? The
method of the topological bootstrap involves coupling two systems
$A$ and $B$, where $A$ is the system of four Majoranas at each site,
now treated as physical degrees of freedom without any constraints,
and $B$ is a system of free spin-1/2s at each site. The full
Hamiltonian we consider is a combination of the Majorana Hamiltonian
specified above and a Kondo coupling between the Majoranas and the
spins: \beq
H&=&H_A+H_{AB}, \nonumber \\
H_A &=&
-\sum_{\langle j, k \rangle} i \gamma_j \gamma_k, \nonumber \\
H_{AB} &=& g \sum_{l, \alpha, \beta} {\vec
\sigma}_l^{\phantom\dagger} \cdot \left( c_{l, \alpha}^\dagger \vec
\tau_{\alpha \beta}^{\phantom\dagger} c_{l, \beta}^{\phantom\dagger}
\right), \nonumber \eeq where $\vec \tau_{\alpha \beta}$ is a vector
of the Pauli matrices, and the four Majorana fermions
$\gamma^{1,2,3,4}_l$ around any site $l$ form a spinful complex
fermion given by \beq && c_{l, \uparrow} = \frac{\gamma^1_l + i
\gamma^2_l}{2},~~c_{l, \downarrow} = \frac{\gamma^3_l + i
\gamma^4_l}{2}. \nonumber \eeq As noted in
Ref.~\onlinecite{bootstrap}, the desired parent Hamiltonian is the
lowest-order effective Hamiltonian for the $B$ system that can be
obtained from degenerate perturbation theory. In this case, the
lowest-order interaction preserving the Majorana integrals of motion
is generated in perturbation theory by four applications of the
Kondo interaction and it is precisely the four-spin interaction of
the Wen plaquette model [see Fig.~6(a) for the Majoranas involved].

Thus, the heuristic is to find the minimal physical interaction,
which, when written in terms of the partons, is a product of parton
integrals of motion. One can then readily check that such physical
interactions do indeed give rise to a suitable parent Hamiltonian.
For example, the four-spin interaction of the Wen plaquette model is
equivalent to the eight-Majorana interaction shown in Fig.~6(a).
Since the link bilinears are each unity in the Majorana ground
state, the product of all eight Majoranas is also unity, and thus
the (projected) Majorana ground state is indeed the ground state of
the Wen plaquette model.

In the same way, we can apply this heuristic to obtain the parent
Hamiltonians described in the main text. In Fig.~6(b,c,d), we
explicitly illustrate the physical interactions of our three fracton
models in terms of the Majorana partons. In each case, one can check
that the resulting product of Majoranas is unity for the Majorana
ground state.

\begin{figure}
\centering
\includegraphics[height=3.0in]{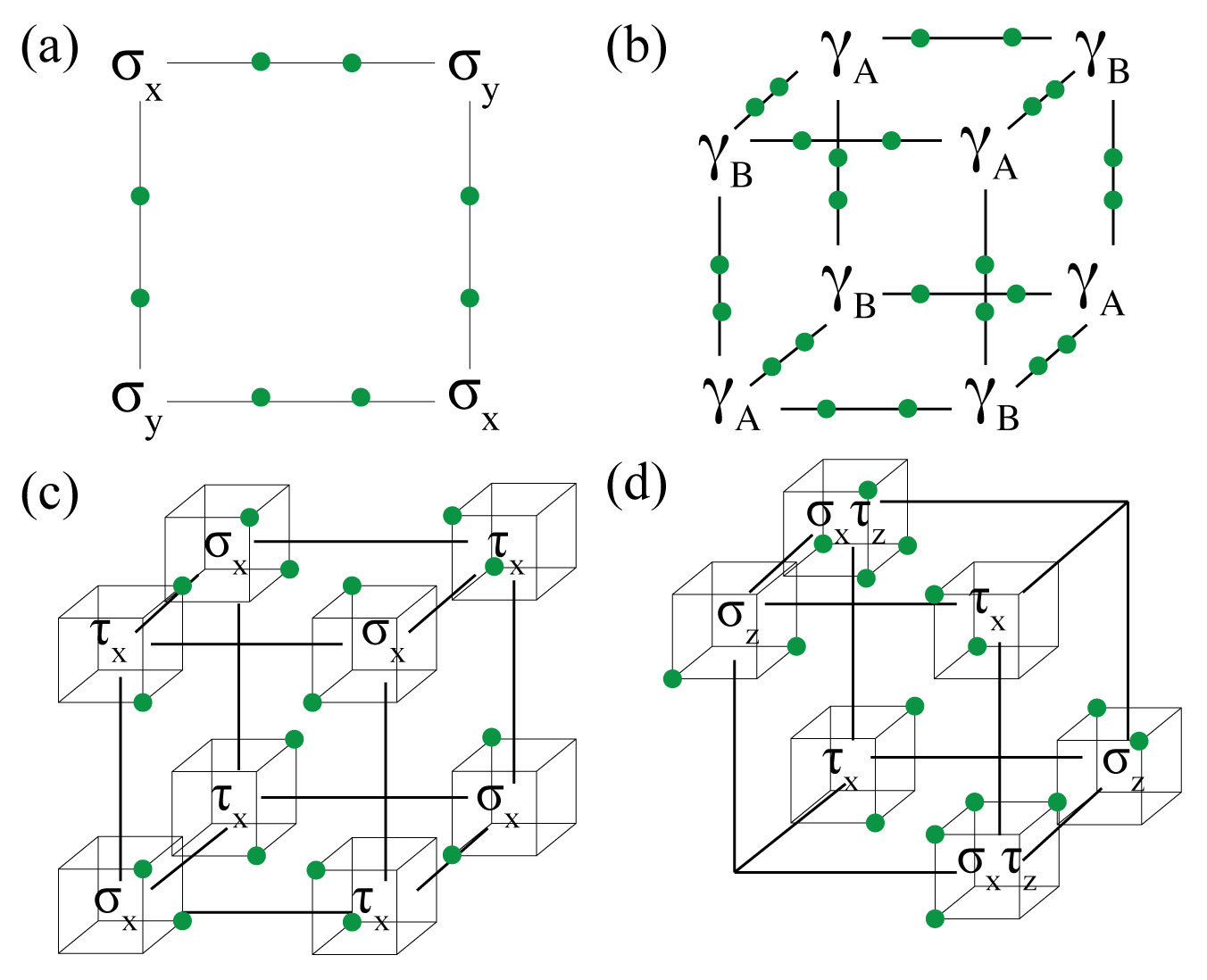}
\caption{Physical interactions in terms of the Majorana partons
(green dots) for (a) the Wen plaquette model, (b) the fermionic
fracton model, (c) the bosonic type-I fracton model, and (d) the
bosonic type-II fracton model. In (c) and (d), only one interaction
is shown; the other interactions are related by simultaneous
threefold rotations in real space and spin space.}
\end{figure}

\clearpage

\end{widetext}


\begin{thebibliography}{10}
\bibitem{footnote-1}
Note that each interaction term is inversion symmetric while the two
independent terms (and their dependent product) are transformed into
each other by a simultaneous three-fold rotation around the [111]
diagonal in both real space and spin space.
\end{thebibliography}
\end{document}